**Nitrospinics as a platform from orbital-torque memory to artificial intelligence**

*Prabhat Kumar [1], Gaurav Kumar Shukla [1], Yoshio Miura [1,2], and Shinji Isogami [1]**

[1] Research Center for Magnetic and Spintronic Materials, National Institute for Materials Science, Tsukuba, JAPAN

[2] Faculty of Electrical Engineering and Electronics, Kyoto Institute of Technology. Hashikami-cho, Matsugasaki, Sakyo-ku, Kyoto, JAPAN

*E-mail: isogami.shinji@nims.go.jp

Keywords: Nitrospinics, Orbital torques, Spin-orbit torques, 2D-MXene, Artificial intelligence

**Abstract**

The exploration of energy-efficient and functional spintronics has attracted considerable attention. Orbital transport has opened new pathways for current-induced torque generation beyond the conventional spin transport based on the spin Hall effect. In addition, artificial intelligence computing has been demonstrated using spintronic devices. Further progress of these devices can be anticipated through the development of unique and functional materials, beyond the existing heavy metals and topological systems with strong spin–orbit coupling. The nitride materials exhibit unique chemical, magnetic, and structural versatility, including antiferromagnetism, high thermal stability, and compatibility with diverse device architectures. Here, we propose Nitrospinics as a conceptual and functional framework that exploits nitride materials for applications ranging from orbital-torque-based spintronic devices to artificial intelligence hardware. Using $Cr_2N$, a two-dimensional nitride MXene with an atomic layered structure, as a prototype system, we discuss how nitrogen contributes to the structural stability, the orbital torque generation, and the interfacial orbital and spin conversion. We further outline key challenges and opportunities toward establishing nitride-based materials for next-generation computing technologies.

## 1. Introduction:

Various phenomena and functionalities driven by spin currents, which correspond to the flow of electron spins, have been increasingly clarified. Electrons possess two degrees of freedom, namely, spin angular momentum and orbital angular momentum, and studies on nonequilibrium transport phenomena have conventionally focused mainly on the spin degree of freedom, namely spin currents, rather than orbital currents. It is widely recognized that spin currents originate from charge-to-spin conversion mechanisms mediated by spin–orbit interaction (SOI), such as the spin Hall effect [1] and the Rashba effect [2].

In recent years, on the other hand, increasing attention has been directed toward the orbital degree of freedom, and fundamental studies aimed at elucidating the origin of orbital currents and establishing methods for their control have been actively conducted. For example, current-induced torques observed in surface-oxidized Cu thin films [3,4] and at $Cu/Al_2O_3$ interfaces [5] have been reported to originate from orbital torques induced by the orbital Rashba–Edelstein effect (OREE) [3–5]. In addition, the phenomenon, in which an orbital current is generated in a direction transverse to an applied electric field, is referred to as the orbital Hall effect (OHE) [6]. Since both the OREE and the OHE can emerge without relying directly on strong SOI, they are attracting considerable interest as novel transport phenomena that can potentially be realized in a wide range of material systems, such as 3*d*, 4*d*, and 5*d* transition metals. For example, the spin Hall conductivities ($\sigma_{\mathrm{SH}}$) of Ti, V, and Cr are generally small because of their negligible SOI, whereas sizable orbital Hall conductivities ($\sigma_{\mathrm{OH}}$) have been reported in the range of $10^3$~$10^4$ $(\hbar/e)$ S/cm [7]. It is well known that the $\sigma_{\mathrm{SH}}$ of Pt is remarkably large, reaching ~$10^3$ $(\hbar/e)$ S/cm, and the $\sigma_{\mathrm{OH}}$ also exhibits the same order of magnitude [7].

In ferromagnetic (FM) / nonmagnetic (NM) bilayers, both channels of $\sigma_{\mathrm{SH}}$ and $\sigma_{\mathrm{OH}}$ can contribute to the effective spin-Hall angle ($\theta_{\mathrm{SH}}^{\mathrm{eff}}$) as a linear combination [8],

$$\theta_{\mathrm{SH}}^{\mathrm{eff}} = (2e/\hbar)(\sigma_{\mathrm{SH}}^{\mathrm{NM}} + \sigma_{\mathrm{OH}}^{\mathrm{NM}}\eta_{L-S}^{\mathrm{FM}})/\sigma_{xx}^{\mathrm{NM}} \,, \tag{1}$$

where $\eta_{L-S}^{\mathrm{FM}}$ and $\sigma_{xx}^{\mathrm{NM}}$ are the conversion coefficient from orbital angular momentum (*L*) to spin angular momentum (*S*) in an FM layer on an NM layer and the electrical conductivity of a NM layer, respectively. Therefore, not only the presence of a large $\sigma_{\mathrm{SH}}$ but also a large $\sigma_{\mathrm{OH}}$ can exert efficient torques by employing the appropriate $\eta_{L-S}^{\mathrm{FM}}$, so that the formula with two terms satisfy $\left(\sigma_{\mathrm{SH}}^{\mathrm{NM}} \cdot \sigma_{\mathrm{OH}}^{\mathrm{NM}}\eta_{L-S}^{\mathrm{FM}}\right) > 0$. For example, the $\sigma_{\mathrm{SH}}$ and $\sigma_{\mathrm{OH}}$ of Pt act synergistically (antagonistically) when the sign of $\eta_{L-S}^{\mathrm{FM}}$ is positive (negative) because both $\sigma_{\mathrm{SH}}$ and $\sigma_{\mathrm{OH}}$ are positive, while the $\sigma_{\mathrm{SH}}$ and $\sigma_{\mathrm{OH}}$ of W and Ta act synergistically (antagonistically) when the sign of $\eta_{L-S}^{\mathrm{FM}}$ is negative (positive) because the sign of $\sigma_{\mathrm{SH}}$ and $\sigma_{\mathrm{OH}}$ is respectively negative and positive. That is, the enhanced $\theta_{\mathrm{SH}}^{\mathrm{eff}}$ is strongly dependent on the sign and magnitude of $\eta_{L-S}^{\mathrm{FM}}$. Given that the Co, Fe, Ni and their alloys are commonly employed as representative FMs in bilayers, the material design space is not enough for the bilayers to achieve dramatic enhancement in the $\theta_{\mathrm{SH}}^{\mathrm{eff}}$. Furthermore, some heavy metals are often expensive and environmentally burdensome. Conversely, the number of transition-metal elements is much larger than that of heavy-metal elements, namely, the wide range of elements and their alloys may provide remarkably high $\sigma_{\mathrm{OH}}$. Not only that, sustainability and harmlessness are the positive characteristics for transition metals, which may be valuable for manufacture. Therefore, it is considered that the 3$d$, 4$d$ and 5$d$ transition mentals have potential to serve as additional spin sources owing to their high $\sigma_{\mathrm{OH}}$, resulting in the significant enhancement of $\theta_{\mathrm{SH}}^{\mathrm{eff}}$ exceeding the values in Pt ($\theta_{\mathrm{SH}}^{\mathrm{eff}} \approx 0.05 \sim 0.1$), Ta ($\theta_{\mathrm{SH}}^{\mathrm{eff}} \approx -0.1 \sim -0.2$), and W ($\theta_{\mathrm{SH}}^{\mathrm{eff}} \approx -0.2 \sim -0.4$).

The next generation of electronic devices is expected to involve two-dimensional (2D) layered materials, such as graphene and transition-metal dichalcogenides (TMDCs), owing to their scalability and compatibility with existing integrated circuits. In particular, for energy-efficient spintronic devices, magnetic tunnel junctions (MTJs) and/or spin-orbit torque magnetic random-access memories (SOT-MRAMs) incorporating 2D materials have attracted significant attention because they are expected to be embedded in the CMOS devices with similar 2D materials in the future [9-11]. Therefore, development of spintronic phenomena in new 2D layered materials is indispensable as well as the transition metals.

In this perspective article, we have addressed a new class of 2D layered material, MXene ($M_{\mathrm{n+1}}X_{\mathrm{n}}$), where $M$ and $X$ represent transition metals and carbon or nitrogen, respectively. Then we further discuss future perspectives extending from orbital-torque memory devices to artificial intelligence hardware. The $Cr_2N$ 2D-layered MXene is epitaxially grown via a reactive nitride sputtering process, and field-free current-induced magnetization switching (CIMS) has been successfully demonstrated in the $Cr_2N$/FM bilayers. Note that the $Cr_2N$ consists of 3$d$ transition metals and nitrogen despite its negligible SOI; therefore, the spin-current might play only a minor role in the possible switching mechanisms, while orbital currents are considered as a possible origin of the observed CIMS [12]. These results are considered specific characteristics realized by the nitride-based layered materials that cannot be achieved using existing SOT materials. Nitrogen has a great impact on the electronic states of transition metals, resulting in sizable $\sigma_{\mathrm{SH}}$ and/or $\sigma_{\mathrm{OH}}$. Although the nitride materials exhibit rich chemical tunability, diverse magnetic ground states, high thermal stability, and compatibility with low-dimensional architectures, they have not yet been widely explored in terms of orbital spintronics.

These perspectives lead to create a new material platform "Nitrospinics," for efficient orbital-torque spintronics based on the material development with nitrogen and facilitate the next-generation artificial intelligence hardware by, e.g., nonvolatile memories, as illustrated in Fig. 1.

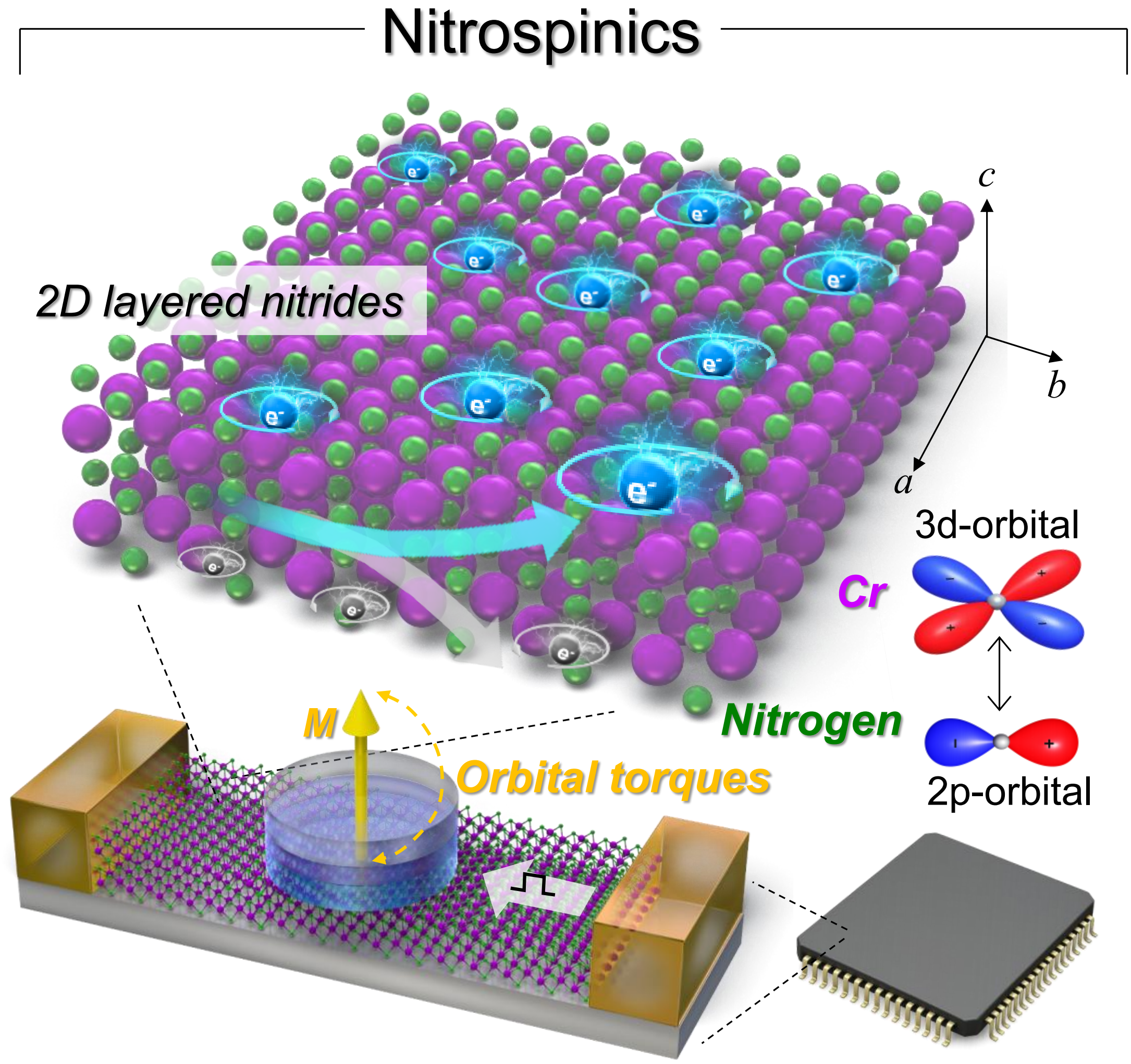

**Fig. 1** Schematic illustration of the Nitrospinics concept, in which two-dimensional (2D) layered nitrides, such as MXenes and/or transition-metal nitrides, serve as a materials platform for orbital-torque-based nonvolatile magnetic memories. The key characteristics are as follows: (i) the memory devices are composed of sustainable materials; (ii) orbital torques are the dominant mechanism responsible for current-induced magnetization switching (CIMS); (iii) field-free CIMS of perpendicular magnetization is enabled by the breaking of local crystal symmetry in nitrides; (iv) robust operation at elevated temperatures; and (v) enhanced torque efficiency compared with pure transition metals, arising from strong orbital hybridization between the $3d$ orbitals of transition metals and the $2p$ orbitals of nitrogen. These characteristics are not achievable with conventional spintronic materials based on heavy metals and their alloys.

## 2. What is Nitrospinics?

Within the context of functionalities enabled by nitride-based materials, transition-metal nitrides have exhibited a variety of intriguing spintronic phenomena, such as inverse tunneling magnetoresistance (TMR) effect [13], inverse current-induced magnetization switching (CIMS) [14], negative anisotropic magnetoresistance [15], giant spin-pumping effects [16], and the topological Hall effect [17]. Regarding spintronic devices, low-power spin logic circuits comprising two types of MTJs with inverse and conventional TMR characteristics have been demonstrated [18], owing to the negative spin polarization of $Fe_4N$ electrodes [19,20]. In addition, SOT-MRAM devices based on transition-metal-nitride/ferromagnet bilayers have attracted considerable attention because of their low power consumption and high efficiency [21–25], while conventional W/CoFeB bilayers have served as the standard platform for SOT-MRAM [26]. These results indicate that nitrogen plays an indispensable role in expanding the functionalities of future devices [27].

However, the potential of nitride materials for SOT-MRAM applications remains insufficiently explored, particularly for the field-free CIMS driven by orbital torques originating from the 2D layered nitrides as depicted in Fig. 1. Nitrospinics is defined as a materials and device framework that utilizes the chemical and magnetic versatility of nitrides to enable spin- and orbital-current generation, torque functionalities, and energy-efficient device operation. Therefore, 2D layered nitrides represent one of the most promising material platforms for realizing the Nitrospinics concept.

### 2.1. Crystal symmetry with interstitial nitrogen atoms

Figure 2 shows the unit-cell model of antiperovskite nitrides ($X_4$N) (space group: $Pm\bar{3}m$), which are among the stable compounds composed of transition metals ($X$) and nitrogen (N), where $X$ atoms form a face-centered cubic (FCC) structure. The stoichiometric $X_4$N contains one N atom at the octahedral center surrounded by six face-centered $X$ atoms, while the off-stoichiometric $X_4$N contains excess and/or deficient N atoms occupying two possible sites: the equivalent octahedral-center sites (blue circles) [Wyckoff position (1/2, 0, 0)] and the metastable sites (red circles) [Wyckoff position (1/4, 1/4, 1/4)]. The FCC structure is preserved regardless of the N occupation sites, which can be attributed to the relatively small atomic radius of N, approximately half that of the $X$ atoms. N is generally regarded as an interstitial impurity and acts to break the local symmetry of the antiperovskite crystal structure. The crystal symmetry ranges from long-range order in the stoichiometric $X_4$N to short-range order in the off-stoichiometric phases, leading to a wide variety of spin and orbital transport properties, without changing the bone crystal structure of $X$. This represents a unique characteristic in the antiperovskite nitrides that cannot be realized in conventional alloy systems.

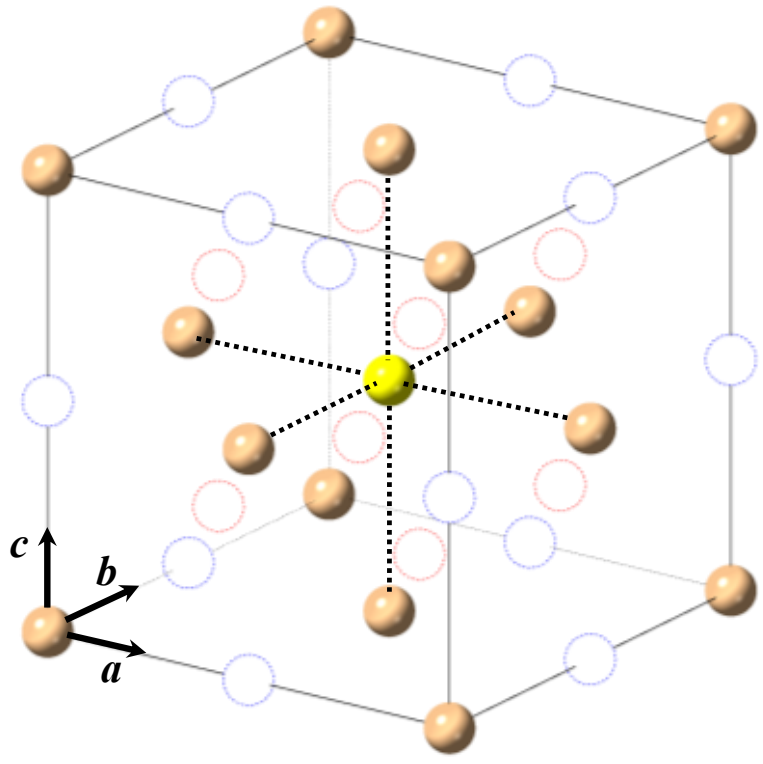

**Fig. 2** Unit-cell model of antiperovskite nitrides $X_4$N, where the $X$ sites are predominantly occupied by transition-metal elements such as Fe, Co, and Mn. The orange and yellow spheres represent $X$ and N atoms, respectively. In off-stoichiometric $X_4$N, excess and/or deficient nitrogen atoms can occupy equivalent sites (blue spheres) [Wyckoff position (1/2, 0, 0)] and/or metastable sites (red spheres) [Wyckoff position (1/4, 1/4, 1/4)] in addition to the octahedral-center site (yellow sphere). This unique feature is enabled by the interstitial nature of nitrogen within the face-centered cubic (FCC) structure.

**2.2. Band characteristics modulated by nitrogen**

In addition to local symmetry breaking induced by N in antiperovskite crystals, unique electronic states due to the hybridization between the $d$ orbitals of $X$ and the $p$ orbitals of N can generally be expected. Such states cannot be realized in conventional alloy systems, where only $d$–$d$ hybridization is involved. To clarify the impact of N on the electronic states, first-principles calculations were performed for stoichiometric antiperovskite crystals with and without N atoms. Figure 3(a) (left) shows the unit-cell of $L1_2$-ordered $Mn_3Pt$, which is a cubic $Cu_3Au$-type compound with a stable noncoplanar antiferromagnetic structure. The Néel temperature ($T_N$) has been reported to be approximately 475 K [28,29]. Figure 3(a) (right) shows antiperovskite $Mn_3PtN$, which also exhibits a stable noncoplanar antiferromagnetic structure. The crystal structure is cubic with a lattice constant of approximately 0.39 nm, comparable to that of $Mn_3Pt$. The only structural difference is the presence of a N atom at the octahedral center of $Mn_3Pt$, which allows us to extract the impact of N by comparison.

Figure 3(b) shows the spin Berry curvature, $\boldsymbol{\Omega}_{xz}^{y}(\boldsymbol{k}, \boldsymbol{E_F})$, at the Fermi energy ($E_F$) along the high-symmetry lines of $Mn_3Pt$ and $Mn_3PtN$. It was revealed that $\boldsymbol{\Omega}_{xz}^{y}(\boldsymbol{k}, \boldsymbol{E_F})$ along the A–Z–R line of $Mn_3PtN$ is larger than that of $Mn_3Pt$, which can be attributed to hybridization between the $d$ orbitals of Mn and the $p$ orbitals of N [27]. To investigate this hybridization effect, we performed a projection analysis of the atomic-orbital contributions to the band dispersions along the high-symmetry lines, as shown in Figs. 3(c) and 3(d). It was revealed that the bands near the $E_F$ in $Mn_3Pt$ were mainly composed of the *d* orbitals of Mn and Pt, while those in $Mn_3PtN$ contained substantial contributions from Mn and nitrogen orbitals, particularly along the A–Z–R line. In addition, a dominant contribution from nitrogen orbitals was observed around −1.0 eV. These bands were considered to be primarily responsible for the enhanced $\boldsymbol{\Omega}^{\boldsymbol{y}}_{\boldsymbol{xz}}(\boldsymbol{k}, \boldsymbol{E_F})$ in $Mn_3PtN$ compared with $Mn_3Pt$, owing to the *p*-*d* dipole transition in $Mn_3PtN$.

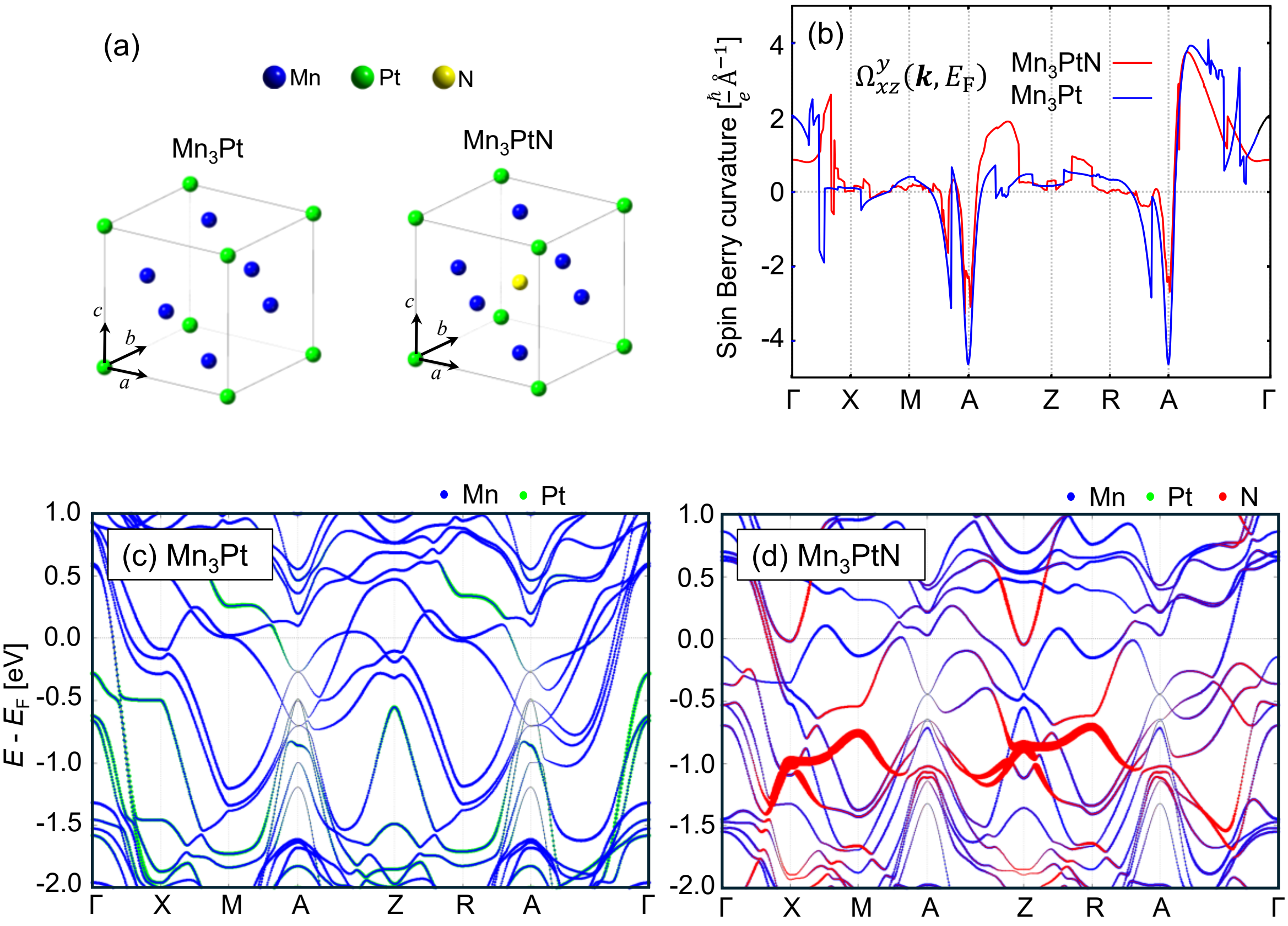

**Fig. 3** (a) Unit-cell models of $Mn_3Pt$ and $Mn_3PtN$. (b) Spin Berry curvature, $\boldsymbol{\Omega}^{\boldsymbol{y}}_{\boldsymbol{xz}}(\boldsymbol{k}, \boldsymbol{E_F})$, at the Fermi energy ($E_F$) along the high-symmetry lines in the first Brillouin zone for $Mn_3Pt$ (red) and $Mn_3PtN$ (blue). (c,d) Projections of the atomic-orbital contributions to the band dispersions of $Mn_3Pt$ and $Mn_3PtN$ along the high-symmetry lines near $E_F$. The projections of the nitrogen orbitals are magnified by a factor of three relative to those of the other atomic orbitals. The high-symmetry *k*-points $(x, y, z)$ are Γ(0, 0, 0), X(0, 1/2, 0), M(1/2, 1/2, 0), A(1/2, 1/2, 1/2), Z(0, 0, 1/2), and R(0, 1/2, 1/2) in the Brillouin zone defined by $k = xb_1 + yb_2 + zb_3$, respectively, where $b_1$, $b_2$, and $b_3$ are the reciprocal lattice vectors of the tetragonal cell. Reproduced with permission from Ref. [25]. Copyright 2024, American Physical Society.

## 3. Prototype of Nitrospinics materials for orbital torque devices

### 3.1. $Cr_2N$ 2D-MXene and formability

One of the most intriguing phenomena realized in Nitrospinics materials is the generation of spin torques originating from orbital currents. As mentioned in the Introduction, orbital currents are generally generated by two major mechanisms: the OREE and the OHE. We have investigated orbital torques originating from the OHE in MXenes, a recently emerging class of 2D layered materials similar to the graphene and TMDC.

The chemical formula of MXenes is $M_{n+1}X_nT_x$, where $M$, $X$, and $T$ represent transition metals such as Ti, V, and Cr; light elements such as N or C; and surface terminations such as Cl, O, and OH, respectively. The parameter $n$ typically varies from 1 to 4, while $x$ is variable. The absence of surface terminations ($x = 0$) is also allowed, which is referred to as a bare MXene. The chemical properties can be tailored by $n$, $x$, and the constituent elements, owing to the strong electronegativity of N or C in the $M$–$X$–$M$ bonding network. Therefore, the electronic and magnetic properties of bare MXenes, consisting of the thinnest $M$–$X$–$M$ trilayers, have been extensively investigated for various phases [30]. In addition, MXenes have been widely studied for applications in biomedicine [31], mechanical engineering [32], optoelectronics [33], and energy storage [34] over the past decade.

$Cr_2N$ is a unique MXene that can be grown using a conventional reactive nitride sputtering technique, the feasibility of which is shown in Fig. 4. The unit-cell exhibits a hexagonal crystal structure [Fig. 4(a1)]. The Cr magnetic moments are indicated by blue and green arrows, suggesting a collinear antiferromagnetic structure [35]. Figures 4(a2) and 4(a3) depict the in-plane and out-of-plane crystal structures of $Cr_2N$ with top and bottom N terminations on the outer Cr layers, which is similar to the structure of TMDCs.

Figure 4(b1) shows the out-of-plane X-ray diffraction (XRD) profiles of 20-nm-thick pure Cr, $Cr_2N$ MXene, and CrN films deposited on $c$-plane-oriented $Al_2O_3$ substrates. Note that CrN, which possesses a cubic NaCl-type structure (space group: $Fm\bar{3}m$), is another stable phase in the Cr–N system. However, we confirmed that single phase $Cr_2N$ MXene without CrN contamination can be obtained by carefully tuning the nitrogen gas flow rate, which may be attributed to the large difference in formation energy between $Cr_2N$ and CrN.

Figure 4(b2) shows the substrate-temperature dependence of the XRD profiles. It was revealed that N does not desorb from the $Cr_2N$ phase up to approximately 650 °C, indicating remarkable thermal stability. Such high stability is highly advantageous for realizing orbital-torque devices, particularly in terms of robustness during device fabrication and circuit integration.

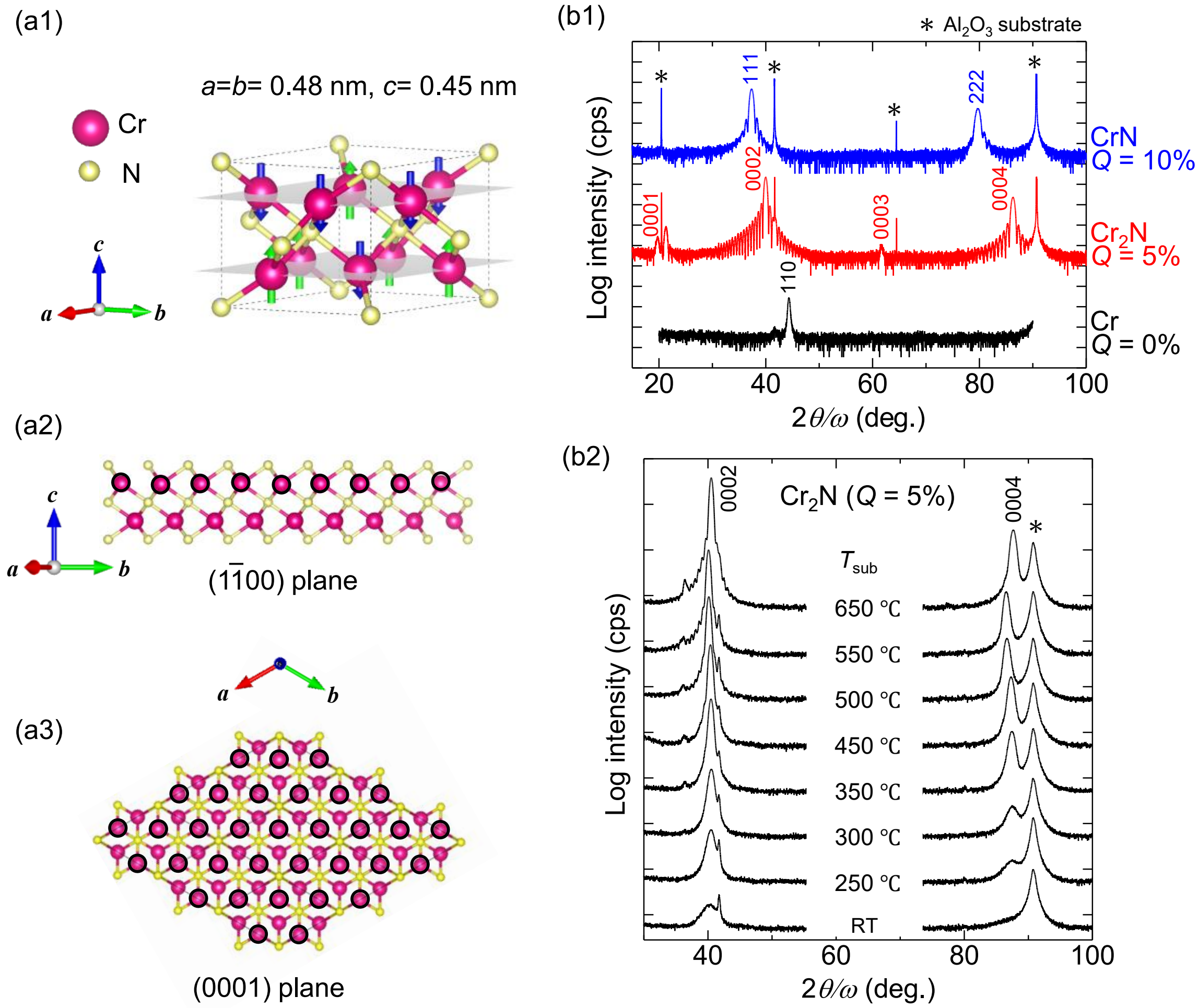

**Fig. 4** Schematic illustrations of the $Cr_2N$ unit-cell with a hexagonal crystal structure (a1), the (1$\bar{1}$00) plane of a single $Cr_2N$ layer terminated by N atoms (a2), and the (0001) plane (a3), where the top Cr layer is highlighted by black circles for clarity. (b) Out-of-plane X-ray diffraction (XRD) profiles of 20-nm-thick pure Cr, $Cr_2N$, and CrN films deposited on *c*-plane-oriented $Al_2O_3$ substrates (b1), and the growth-temperature dependence of the crystal structure of 20-nm-thick $Cr_2N$ films (b2). Reproduced from Ref. [24] under CC-BY-4.0.

### 3.2. Observation of current-induced field-free magnetization switching

CIMS has been demonstrated in the *c*-plane-oriented $Al_2O_3$ substrate // $Cr_2N$ (5 nm) / [Co (0.35 nm)/Pt (0.3 nm)]$_3$ / MgO (3 nm) structures for two different current-pulse (*I*) directions with respect to the in-plane mirror-symmetry line (*m*), as shown in Fig. 5, namely, ($m \perp I$) and ($m \parallel I$). Microfabrication was performed to form a dot-patterned [Co/Pt]$_3$/MgO layer on top of the $Cr_2N$ channel layer, as shown in the polar Kerr microscopy images.

The magnetic easy-axis was oriented along the out-of-plane direction, as evidenced by the out-of-plane anomalous Hall hysteresis loops [Figs. 5(a2) and 5(b2)]. Deterministic CIMS was observed under an applied in-plane magnetic field, as confirmed by the CIMS loops [Figs.

5(a3) and 5(b3)]. Notably, field-free CIMS was also achieved regardless of two current-pulse directions, as shown in Figs. 5(a4) and 5(b4).

For the CIMS of perpendicular magnetization driven by the spin current with spin polarization along the *y* direction (spin-Y), an in-plane magnetic field must be applied along the *x* direction, parallel to the charge-current direction, to break the inversion symmetry. This mechanism is referred to as the conventional SOT and is typically observed in bilayer structures comprising FM layer and heavy metal layer such as W, Pt, and Ta.

In contrast, several approaches have been proposed to achieve field-free CIMS, which are thickness and compositional gradients [36], magnetization canting states [37], and hybrid devices combining SOT and spin-transfer torque (STT) [38]. Furthermore, the injection of spin polarization along the *z* direction (spin-Z) induced by crystal symmetry has also been proposed as a possible mechanism for field-free CIMS [39]. These approaches are generally classified as unconventional SOT mechanisms.

The role of crystal symmetry in generating spin-Z is illustrated using the crystal structure of 2D $WTe_2$, which exhibits broken symmetry [40]. It has been revealed that spin-Z emerges when the charge current flows along the low-symmetry direction ($m \perp I$), while only spin-Y is allowed when the charge current flows along the high-symmetry direction ($m \parallel I$). These characteristics give rise to anisotropic switching properties, particularly in devices based on 2D materials, and have been regarded as a challenge for practical device implementation.

Notably, such anisotropic switching property was not observed in devices based on 2D MXenes, as shown in Fig. 5. This behavior may be attributed to an isotropic switching mechanism distinct from crystal-symmetry-driven effects, such as uncompensated interfacial magnetic moments induced by the adjacent ferromagnetic layer [41]. This represents one of the unique characteristics of $Cr_2N$-MXene-based devices that cannot be readily realized in devices employing TMDC. Therefore, the mechanisms distinct from those operating in conventional TMDC systems are considered responsible for the field-free CIMS observed in $Cr_2N$.

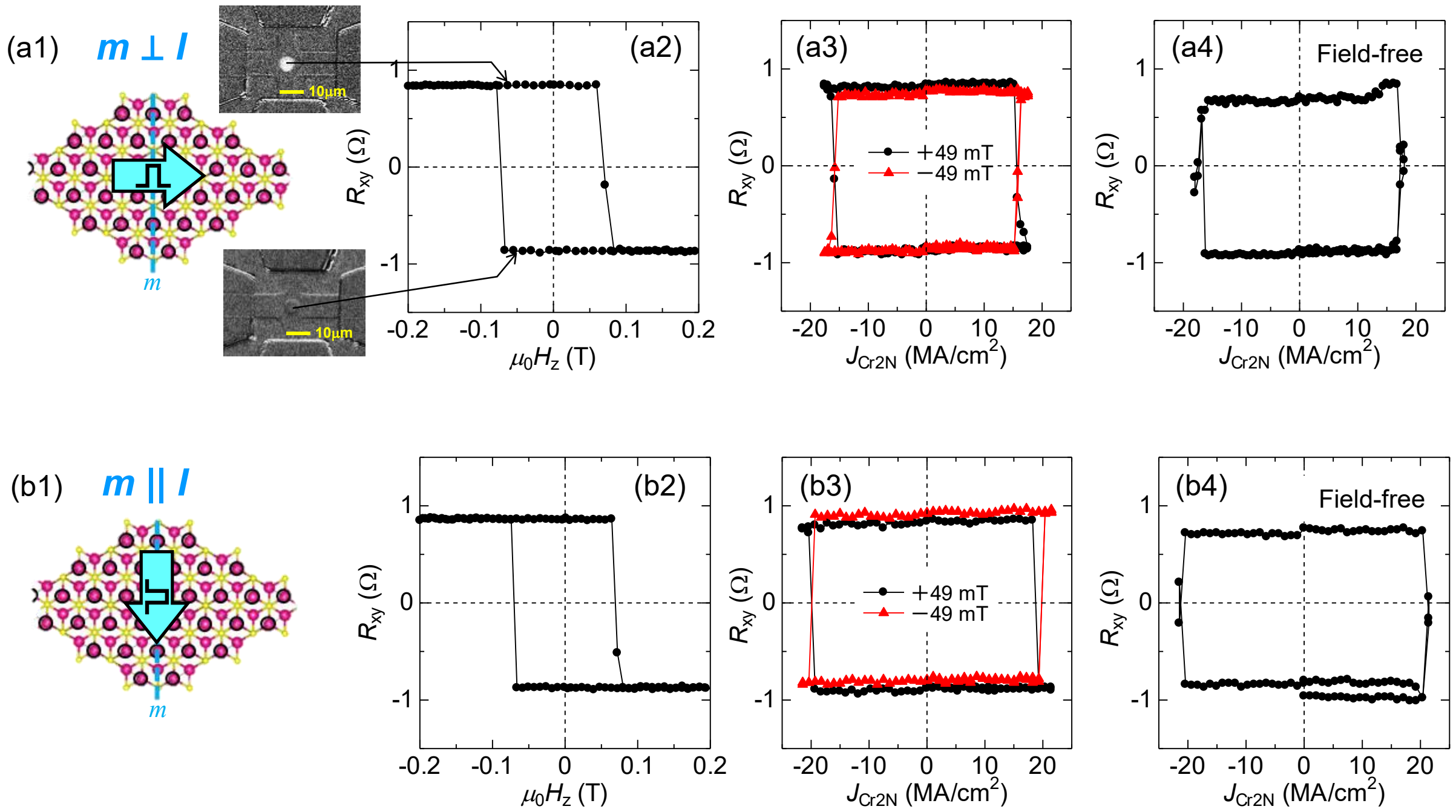

**Fig. 5** (a) Crystal structure of the $Cr_2N$ (0001) plane together with the current-pulse ($I$) direction with respect to the in-plane mirror symmetry line ($m$) (a1). Out-of-plane anomalous Hall effect (AHE) hysteresis loops (a2), and current-induced magnetization switching (CIMS) measured with (a3) and without (a4) an in-plane magnetic field for the stack structure: $c$-$Al_2O_3$ substrate // $Cr_2N$ (5 nm) / [Co (0.35 nm)/Pt (0.3 nm)]$_3$ / cap, where the $I$ flows orthogonal to the $m$ line ($m \perp I$). The insets show polar Kerr microscopy images of the microfabricated pillar containing the [Co (0.35 nm)/Pt (0.3 nm)]$_3$/cap structure. (b) Same data as Fig. (a) but for the case of ($m$ // $I$).

## 4. Functional opportunities by Nitrospinics

### 4.1. Robustness, stability and scalability in nonvolatile memories

Table I summarizes the properties of nonvolatile magnetic memory devices comprising orbital source layers and FM layers, which are typically reported to date. Various pure transition metals, such as Ti and Cr, have been employed as orbital source layers, while a wide variety of alloys and even 2D materials have been used as FM layers. Although these materials are not heavy metals with strong SOI, their torque efficiencies are comparable to those of the conventional $\beta$-W/CoFeB system (−0.33 to −0.40) [42,43]. Furthermore, CIMS has been demonstrated using current pulses with amplitudes and pulse durations comparable to those employed in $\beta$-W/CoFeB systems. The possible mechanism responsible for these CIMS behaviors is the spin torque originating from orbital currents generated by the OHE and subsequent orbital-to-spin conversion, commonly referred to as orbital torque. This mechanism is fundamentally different from conventional spin–orbit torques generated through the spin Hall effect (SHE). Therefore, the materials listed in Table I contribute to expanding the range of torque sources beyond the currently utilized heavy metals.

Nevertheless, field-free CIMS with reduced critical current density ($J_c$) remains highly desirable for robust and scalable nonvolatile memory applications. One possible route to realize such functionality is the incorporation of N into the orbital source layers. Transition-metal nitrides such as CrN and VN have been reported to exhibit different torque efficiencies in bilayer structures compared with their pure metal counterparts, Cr and V. The variation in torque efficiency observed in CrN- and VN-based systems may be associated with differences in layer thickness, suggesting that the OHE induced torque efficiency can be widely tuned through thickness engineering. On the other hand, bilayer structures consisting of $Cr_2N$ MXene and Co/Pt ferromagnetic multilayers exhibit field-free CIMS, a phenomenon that has not been reported in other orbital source materials.

It should be noted that nitrides as OHE source materials are not expected to replace existing SOT devices immediately, as several challenges remain before practical applications such as MRAM can be realized, including further reduction of $J_c$ and compatibility with MTJs. Nevertheless, nitrides significantly expand the materials design space for future orbital-torque based devices. For example, enhanced torque efficiency and reduced $J_c$ have been observed in CrN/$[Co/Pt]_3$ compared with pure Cr/Co and Cr/CoFeB systems. In addition, field-free CIMS emerged in $Cr_2N$/$[Co/Pt]_3$, which may be attributed to the reduced crystal symmetry induced by N incorporation. In this context, many nitrides, such as MoN, $Mo_2N$, HfN, and ScN, can be regarded as promising candidate materials for robust and scalable nonvolatile memories. These findings highlight the beneficial role of N in broadening the functional opportunities available for future spintronic devices.

**4.2. Other functionalities**

In addition to robustness, phase stability, and device scalability, the presence of N can modulate the electronic band structure significantly through orbital hybridization between metallic elements and N. For example, the band character of $Fe_4N$ at the $E_F$ has been found to be dominated by the minority-spin states of Fe, as revealed by the direct visualization of the band dispersion using spin and angle resolved photoemission spectroscopy [44]. This indicates that the sign of the spin polarization at $E_F$ becomes negative owing to nitrogen incorporation [45], giving rise to inverse tunneling magnetoresistance (TMR) [46] and negative anisotropic magnetoresistance (AMR) [47]. These unique properties are expected to advance both device engineering and emerging functionalities, such as neuromorphic computing [48,49] and probabilistic devices [50,51], toward future applications in artificial intelligence (AI) hardware and energy-efficient computing.

**Table I** Torque efficiency and critical current density for the reported orbital torque devices

| Orbital source layer (nm) | | FM layer (nm) | Torque efficiency $\xi_{DL}^{eff}$ | $J_c$ (MA/cm$^2$) / field (mT) | Reference |
|---|---|---|---|---|---|
| Ti (50) | Metal | Ni (5) | 0.017 | N.A. | [52] |
| Ti (10) | Metal | $Fe_3GaTe_2$ (16) | N.A. | 3 / 100 | [53] |
| Ti (10)/Pt (2) | Metal | $Fe_3GaTe_2$ (16) | N.A. | 6 / 100 | [53] |
| V (10) | Metal | Co (10) | 0.5 | N.A. | [54] |
| V (10) | Metal | Fe (10) | 0.55 | N.A. | [54] |
| V (10) | Metal | Py (10) | 0.17 | N.A. | [54] |
| Cr (10) | Metal | Co (3) | -0.018 | N.A. | [55] |
| Cr (10) | Metal | Ni (2) | 0.01 | N.A. | [55] |
| Cr (5) | Metal | Gd (10) | -0.21 | N.A. | [55] |
| Cr (10) | Metal | CoFeB (0.9) | -0.02 | 120 / 20 | [55] |
| Cr (10)/Pt (1) | Metal | CoFeB (0.9) | 0.017 | 60 / 20 | [55] |
| Cr (2)/Pt (1.5) | Metal | $[Co(0.2)/Ni(0.6)]_3$ | 0.21 | 100~130 / 50 | [56] |
| Ru (2)/Pt (1.5) | Metal | $[Co(0.2)/Ni(0.6)]_3$ | 0.49 | 80~120 / 50 | [56] |
| Nb (2)/Pt (1.5) | Metal | $[Co(0.2)/Ni(0.6)]_3$ | 0.32 | 140~190 / 50 | [56] |
| Zr (10) | Metal | $[Co(0.3)/Pt(0.7)]_3$ | 0.8 | 3.8 / 20 | [57] |
| $CuO_x$ (6) | Oxide | $Gd_{23}Co_{77}$ (10) | -0.02 | N.A. | [58] |
| CrN (5) | Nitride | $[Co(0.35)/Pt(0.3)]_3$ | 0.09 | 2~2.5 / 20~50 | [23] |
| VN (7.5) | Nitride | $[Co(0.35)/Pt(0.3)]_3$ | -0.41 | 20 / 30 | [59] |
| $Cr_2N$ (5) | 2D-MXene | $[Co(0.35)/Pt(0.3)]_3$ | N.A. | 25~35 / Field free | [24] |

## 5. Application challenges and outlook

### 5.1. Neuromorphic computing devices

AI is a rapidly developing technology that contributes to various sectors of society. However, the power consumption of existing neural network and reservoir computing systems remain significantly higher than that of the human brain. Spintronic technologies have been explored as a potential solution to this challenge owing to their unique advantages, such as low-power consumption, high-speed operation, durability, probabilistic behavior, and compatibility with CMOS technologies. Consequently, various feasibility studies have been conducted to realize edge devices and AI hardware. For example, spintronic neural networks have been demonstrated using MTJs, in which spin-torque nano-oscillators and magnetic skyrmions act as neurons. Reservoir computing has been demonstrated based on the nonlinear magnetization dynamics of magnetic materials [48,49]. Nevertheless, specific materials design is considered indispensable for realizing such systems, because the performance of neuromorphic computing devices is strongly influenced by magnetic anisotropy, SOI, thermal stability, and device scalability. Although nitrides have not yet become major materials for these applications, they have considerable potential as candidate materials owing to their high thermal and phase stability, durability, and the wide tunability of their electronic states through nitrogen engineering.

### 5.2. Probabilistic devices

Since the rapid expansion of AI applications, the limitations of conventional CMOS based computing systems in terms of efficiency and power consumption have become increasingly apparent. In recent years, probabilistic spin logic (PSL) driven by the thermal fluctuations of MTJs has attracted considerable attention, because the output state of a probabilistic bit (p-bit) is controlled between 0 and 1 by external input signals [50]. A 130-nm CMOS-integrated p-bit based on a superparamagnetic MTJ has already been demonstrated [51]. Among the various p-bit architectures, SOT based p-bits are considered particularly attractive, because of their high read/write speed, compatibility with CMOS technology, and the separation of write and read current paths. In such devices, the probabilistic distribution between the 0 and 1 states must be efficiently controlled with low power consumption. However, conventional approaches still face several challenges, including the limited availability of suitable heavy-metal materials and insufficient SOT efficiency. In contrast, significant OHE, out-of-plane orbital torques, and tunable torque efficiencies offered by nitride materials are promising for SOT-based p-bits. These characteristics may provide a new pathway for the development of p-bit technologies based on OHE in nitrides, rather than relying on the SHE in conventional heavy-metal systems.

### 5.3. Thermal-orbitronic computing

As discussed in the context of p-bit devices, large amounts of power are required to control the probabilistic distribution associated with stochastic switching, which is expected to become a critical challenge for future data centers. One promising approach is to utilize self-induced temperature gradients within p-bit devices instead of relying on charge current injection. A possible device architecture is to place p-bit arrays in locations where stable waste heat is naturally generated, such as near CPUs and GPUs. Under such conditions, stochastic switching may be induced through efficient orbital-current injection from the orbital source layer into the p-bit array via orbital Seebeck mechanisms [60]. Therefore, nitride based orbital source materials may contribute significantly to realize a new paradigm for low-power thermal-orbitronic computing through the utilization of waste heat in future data centers.

### 5.4. Ultrafast optical orbital computing

Magneto-optical Kerr effects in nitrides have been investigated to clarify the dependence of magneto-optical inter-band transition probabilities on light-element content [61], suggesting that the content of light elements plays a dominant role in determining magneto-optical properties. Such magneto-optical functionalities may provide a pathway toward optical computing systems that eliminate charge-current flow, thereby enabling p-bit arrays with negligible Joule heating during readout and computation through optical signal transmission along interconnected photonic pathways. Furthermore, the Kerr rotation angle can be tuned resulting from the nitrogen-induced modulation of magnetic ordering and orbital hybridization. This tunability may reach polarization and intensity modulators that act as nonvolatile magneto-optical weights in optical neural networks. Intrinsically high thermal stability of nitrides may allow for ultrafast computation and readout based on femtosecond laser excitation. These developments suggest that magneto-optical computing platforms could be established through tunable Kerr rotation, enabling optical readout, stochastic-bit operation, and reconfigurable photonic neural networks based on nitrides with uniquely tunable magnetic ordering and orbital hybridization.

## 6. Conclusion

We have summarized the future prospects of nitride material systems as a platform for the development of orbital torques. Nitrides generally exhibit fundamental characteristics such as high phase stability, thermal durability, and hardness, which have long been recognized. In addition, unique magnetic and spintronic properties, such as negative spin polarization and enhanced spin torques revealed by recent studies, have attracted considerable attention owing to their potential for diverse device functionalities and materials engineering.

These properties originate from modifications of the electronic band structure originating from orbital hybridization between the *d* orbitals of transition metals and the *p* orbitals of nitrogen. While spin currents generated by the SHE dominate in heavy metals with strong SOI, orbital currents generated by the OHE in transition metals with weak or negligible SOI have recently attracted attention for memory applications driven by sizable orbital torques. Such mechanisms provide an alternative route to the conventional SOT-MRAM technologies.

Devices incorporating nitrides, such as CrN, VN, and even the 2D material $Cr_2N$ MXene, have demonstrated not only enhanced torque efficiencies but also sign reversal of torque and field-free CIMS. These findings indicate that nitrogen in transition metals is highly effective in expanding the range of orbital-torque functionalities. Furthermore, the diverse functionalities of nitrides may be advantageous for realizing neuromorphic devices and probabilistic-bit (p-bit) technologies, originating from the excellent thermal and optical robustness caused by nitride materials. These prospects lead us to conclude that nitrides offer a new pathway from memory technologies to intelligent computing systems, leading to the emerging research field of *Nitrospinics* in the future.

**Acknowledgements**

This work was supported by KAKENHI Grants-in-Aid No. 23K22803 and 25K01275 from the Japan Society for the Promotion of Science (JSPS), Adaptable and Seamless Technology transfer Program through Target-driven R&D (A-STEP) from Japan Science and Technology Agency (JST) No. JPMJTR25T3, Iketani Science and Technology Foundation No. 0371203-A, and Murata Science and Eduction Foundation No. M24AN110. Part of this work was performed under the Cooperative Research Project Program of the RIEC, Tohoku University, and the "X-nics" program of the Ministry of Education, Culture, Sports, Science and Technology (MEXT), Japan.